\documentclass[final,5p,times,twocolumn,authoryear]{elsarticle}
\usepackage{graphicx,amsmath,amssymb}
\usepackage{lipsum}
\usepackage{color}
\usepackage[amssymb]{SIunits}
\usepackage{lscape}

\journal{International Journal of Multiphase Flow}

\begin{document}

\begin{frontmatter}

\title{Salts promote or inhibit bubbly drag reduction in turbulent Taylor\textendash Couette flows}

\author[PoF]{Luuk J. Blaauw}
\author[PoF,MPI]{Detlef Lohse}
\author[PoF]{Sander G. Huisman}
\affiliation[PoF]{organization={Physics of Fluids Department and Max Planck Center for Complex Fluid Dynamics, Department of Science and Technology, and J.M. Burgers Center for Fluid Dynamics, University of Twente},
            addressline={Drienerlolaan 5}, 
            city={Enschede},
            country={The Netherlands}}
\affiliation[MPI]{organization={Max Planck Institute for Dynamics and Self-Organization},
            addressline={Am Fa\ss berg 17}, 
            city={Gottingen},
            country={Germany}}

\begin{abstract}
Bubbly drag reduction is considered as one of the most promising techniques to reduce the energy consumption of marine vessels. With this technique bubbles are injected under the hull where they then lubricate the hull, thus reducing the drag of the vessel. Understanding the effects of salts on bubbly drag reduction is therefore of crucial importance in the application of this technique for salt waters. In this study we investigate the effects of $\mathrm{MgCl_2}$, $\mathrm{Na_2SO_4}$, substitute sea salt, and $\mathrm{NaCH_3COO}$ on the reduction of drag by bubbles in turbulent Taylor--Couette flow. We find that $\mathrm{MgCl_2}$, $\mathrm{Na_2SO_4}$, and substitute sea salt inhibit bubble coalescence, leading to smaller bubbles in the flow, which prove to be less effective for bubbly drag reduction. For these salts we find that the ionic strength is a decent indicator for the observed drag reduction and solutions of these salts with an ionic strength higher than $I\geq\unit{0.7}{mol\per\liter}$ show little to no drag reduction. In contrast, $\mathrm{NaCH_3COO}$ solutions do not inhibit bubble coalescence and for this salt we even observe an enhanced drag reduction with increasing salt concentration. Finally, for all cases we connect the observed drag reduction to the bubble Weber number and show that bubble deformability is of utmost importance for effective bubbly drag reduction.
\end{abstract}


\end{frontmatter}

\section{Introduction}
\label{sec:SaltDR_intro}
The shipping industry emits 1056 million tons of $\mathrm{CO_2}$ yearly (\cite{Faber_2020}). The energy produced by the ship's engine is ultimately mainly eaten up by the drag of the ship. Out of the total drag of a ship, skin friction accounts for 60\% (\cite{larsson_2010_ship}). Reducing the skin friction is therefore crucial in reducing the total drag acting on a ship hull and its concomitant reduction of energy, fuel, and emissions. Bubbly drag reduction is regarded as one of the most promising techniques to reduce the skin friction, which could lead to a significant reduction in the ship's fuel consumption. 

Bubbly drag reduction has received interest for well over a decade and has been extensively reviewed by e.g. \cite{Murai_2014_review,Ceccio_2010_review,Hashim_2015_review}. Different mechanisms for bubbly drag reduction have been proposed to explain the effect, including compressibility (\cite{Lo_2006,ferrante_elghobashi_2004,vdBerg_2005}), bubble coalescence (\cite{Jha_2019,Cannon_2021}), bubble deformability (\cite{Verschoof_2016_surfactant,vGils_2013_bubbledeform,Lu_2005_bubbledeform,vdBerg_2005}), bubbles weakening the large scale structures in the flow (\cite{Murai_2005,murai_2008,spandan_2018}), bubbles pushing vortical structures away from the wall (\cite{ferrante_elghobashi_2004}), and decrease of surface area in contact with water due to bubbles at the surface (\cite{Zverkhovskyi_2014}). Most likely all these mechanisms play \textit{some} role, but when exactly each mechanism is dominant is poorly understood, nor do we know how to exploit them to drastically enhance the drag reduction.

Drag reduction has been investigated in different configurations, such as horizontal (\cite{Jha_2019}) and vertical channel flow (\cite{Lu_2005_bubbledeform}), flat plate experiments (\cite{Madavan_1984,Madavan_1985,elbing_2008,sanders_2006, Kodama_2000}), and Taylor--Couette flow (\cite{Murai_2005,murai_2008,Fokoua_2015,vanBuren_2017}). In flat plate experiments drag reductions of up to 80\% have been reported (\cite{Madavan_1984,Madavan_1985}). In Taylor--Couette flows using an average air volume fraction of 4\% drag reductions of up to 40\% have been achieved (\cite{vGils_2013_bubbledeform}). Real life applications of a bubble injection underneath the ship hull achieved only $\approx10\%$ drag reductions in the best case, and some study even showed a net increase in energy usage, as injecting the bubbles under the hull also costs energy (\cite{Mizokami_2010_mitsubishi,Mizokami_2013_mitsubishi,Kumagai_2015}). Possible explanations could be scaling effects where extrapolations of lab results to real vessel sizes go wrong because these configurations operate in different regimes (\cite{Kodama_2000,Ceccio_2010_review}), contaminations present in seawater (\cite{Blaauw_2023,Jha_2019}), or roughness due to barnacles, welds, or algae growth (\cite{verschoof_2018_roughDR,Bullee_2020_roughness}).

Most lab studies have been performed using purified (fresh) water. Despite that seawater contains a large variety of salts, surfactants, biological matter, and particulate contents, only a few studies have considered water contaminations. \cite{Ma_2023} observed that with an increasing surfactant concentration the bubble induced turbulence increased in strength and the level of anisotropy in the flow is enhanced. \cite{Winkel_2004} experimentally investigated bubble size distributions in a horizontal water channel using fresh water, salt water, and surfactant solutions. They found bubble sizes decrease from approximately $d_{\mathrm{bubble}}\approx\unit{0.4}{\milli\meter}$ for fresh water to $d_{\mathrm{bubble}}\approx\unit{0.1}{\milli\meter}$ in 3.8\% salt solution and $d_{\mathrm{bubble}}\approx\unit{0.2}{\milli\meter}$ in a \unit{20}{ppm} surfactant solution. \cite{shen_2006} found that by injecting bubbles in a $3.5\%$ salt solution and a $\unit{20}{ppm}$ surfactant solution the drag reduction is independent of the bubble sizes. In contrast, \cite{elbing_2008} found that injection of \unit{20}{ppm} surfactant had little to no effect on the mean bubble size in the near wall region and drag reduction. \cite{Verschoof_2016_surfactant} found in bubbly turbulent Taylor--Couette flow with an air volume fraction of 4\% that injection of \unit{6}{ppm} of the surfactant Triton X-100 decreased the bubble sizes and drastically reduced the drag reduction from around 40\% to 4\%.

Salts are known to either inhibit or facilitate bubble coalescence. \cite{Craig_1993} empirically divided anion-cation pairs in two categories where certain combinations inhibited bubble coalescence and others had no effect on bubble coalescence. Multiple explanations have been put forward to explain this coalescence inhibition, such as effect of ionic strength (\cite{Quinn_2014,Sovechles_2015}), electric interactions (\cite{Katsir_2014}), Hofmeister series (\cite{Craig_2004,Henry_2010}), hydration forces (\cite{Tsang_2004}), surface affinity of ions (\cite{weissenborn_1995,Henry_2010}), and effects on surface tension (\cite{Firouzi_2014,Firouzi_2015}), and recently, this effect was explained using Marangoni stresses (\cite{Duignan_2021,Liu_2023}).

In turbulent flows, bubbles come in a range of sizes, i.e. a size distribution which is set by a dynamic equilibrium between bubble break-up and bubble coalescence by collisions, both caused by the underlying turbulent flow. Adding salt to water can inhibit the bubble coalescence leading to a smaller mean equilibrium bubble size in the turbulent flow. Although this coalescence inhibition is caused by effects on the nano-scale, the effects can be significant on macroscopic scales, e.g. for thermal mixing of bubbly flows (\cite{waasdorp_2024}) or bubble shattering (\cite{Slauenwhite_1999}).

Recently, \cite{Biswas_2024} investigated the effect of $\mathrm{MgSO_4}$ on bubbly drag reduction in a horizontal water channel. They saw that at moderate Reynolds numbers $20000\leq\mathrm{Re}\leq45000$ the addition of salt led to smaller bubbles and an increase in drag. At higher Reynolds numbers $45000\leq\mathrm{Re}\leq65000$ the coalescence in fresh water is already hindered by the short collision time, hence the addition of salt only slightly reduced bubble sizes and slightly increased the drag. Moreover, they observed a critical salt concentration after which increasing the salt concentration had no further effect on the bubbles and drag.

In our Taylor--Couette setup (\cite{vGils_2011_T3C}) we previously investigated the effect on $\mathrm{NaCl}$ on bubbly drag reduction (\cite{Blaauw_2023}). We confirmed that sodium chloride inhibits the coalescence, leading to smaller and less deformable bubbles in the flow. These smaller bubbles are less effective for bubbly drag reduction, hence the drag reduction reduces with an increase in salt concentration. For $\mathrm{Re_i} = 2\times10^6$ we observed 40\% drag reduction in fresh water which was reduced to 15\% drag reduction in a $3.5\% (w/w)$ sodium chloride solution. Increasing the salt concentration above the critical coalescence concentration (\cite{Quinn_2014}) does not further inhibit the coalescence, and therefore the bubble sizes and the drag reduction.

In this study we investigate the drag reduction caused by bubbles in Taylor--Couette turbulence. When determining the bubbly drag reduction we already take into account the small changes caused by changes in density and viscosity of the fluid. We investigate the effect of $\mathrm{MgCl_2}$, $\mathrm{Na_2SO_4}$, and synthetic ocean water separately on bubbly drag reduction in Taylor--Couette turbulence, where the first two are the most common salts found in seawater after sodium chloride. Next to that, we investigate the effect of sodium acetate, a salt which should not have any effect on bubble coalescence according to \cite{Craig_1993}.

This manuscript is organised in the following way. In section \ref{sec:SaltDR_exp} we discuss the experimental setup and the salts used in the experiments. Results are discussed in section \ref{sec:SaltDr_Results}. We conclude our research in section \ref{sec:SaltDR_conclusion}.

\section{Experimental methods}
\label{sec:SaltDR_exp}
As in our previous work (\cite{Blaauw_2023}), we utilise the Twente Turbulent Taylor--Couette (T$^3$C) facility (\cite{vGils_2011_T3C}) for our experiments. The Taylor--Couette setup consists of two concentric cylinders that can rotate independently. In the current work only the inner cylinder is rotating, while the outer cylinder, including the top and bottom lids, is kept stationary. The Taylor--Couette setup has a well-defined energy balance, allowing us to carefully evaluate the drag reduction over a large range of rotation velocities. Moreover, the closed volume allows for easy control of the air volume fraction and the salt concentration.

\begin{figure}
    \centering
    \includegraphics[width=0.85\linewidth]{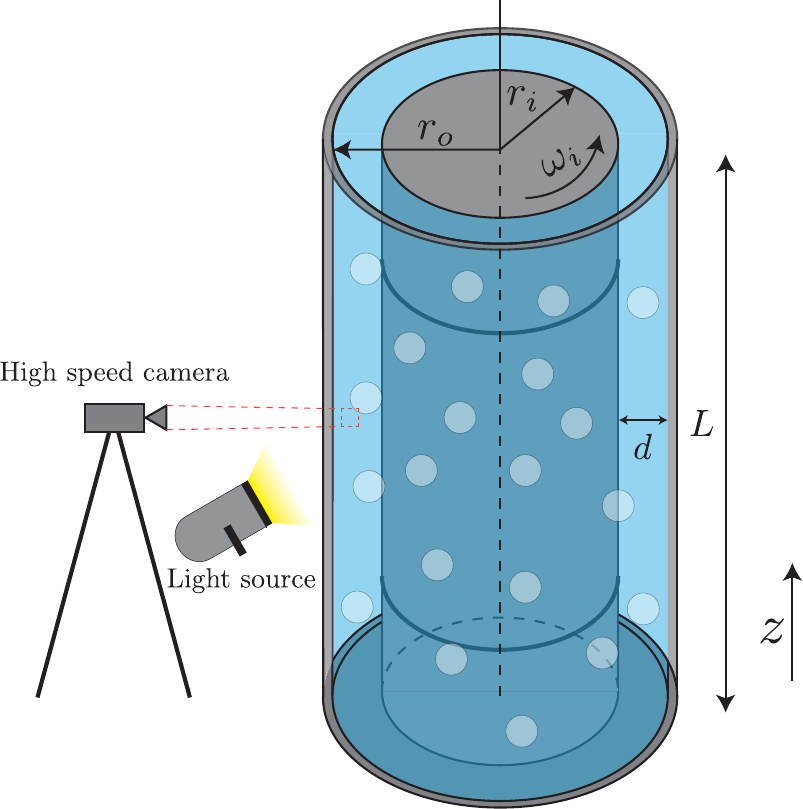}
    \caption{Experimental setup with relevant parameters. The camera on the left hand side has an approximate field of view of $\unit{2.5}{\centi\meter}\times\unit{2.5}{\centi\meter}$. Top lid omitted for visibility. Bubbles are not to scale.}
    \label{fig:saltDR_setup}
\end{figure}

In analogy to the Rayleigh number in Rayleigh--B\'enard convection characterizing the amount of (thermal) driving, we characterise the (mechanical) driving of the Taylor--Couette flow with a Taylor number (\cite{Eckhardt2007b,Grossmann_2016_TCrev}):
\begin{equation}
\label{eq:SaltDR_Taylor}
    \mathrm{Ta} = \frac{(1+\eta)^4}{64\eta^2}\frac{(r_o-r_i)^2(r_i+r_o)^2(\omega_i-\omega_o)^2}{\nu^2},
\end{equation}
where $\eta = r_i/r_o$ is the radius ratio, $r_i=\unit{0.200}{\meter}$ is the inner cylinder radius, $r_o = \unit{0.279}{\meter}$ is the outer cylinder radius, $\omega_{i,o}$ are the angular velocities of the inner and outer cylinder, respectively, and $\nu$ is the kinematic viscosity of the fluid.

The drag is determined by measuring the torque on the inner cylinder, which is axially divided into three parts, see figure \ref{fig:saltDR_setup}. The torque on only the middle part of the inner cylinder is measured to eliminate end effects caused by the stationary lids. The torque is measured using a hollow flanged torque sensor from Althen with a range of \unit{225}{\newton\meter} both ways and is calibrated in situ. We use the angular velocity Nusselt number, analogous to the Nusselt number in Rayleigh--B\'enard convection characterizing the thermal transport, to characterise the angular velocity transport:
\begin{equation}
\label{eq:SaltDR_Nusselt}
    \mathrm{Nu}_\omega = \frac{r_o^2-r_i^2}{4\pi L_\mathrm{mid} \nu\rho (\omega_i-\omega_o)r_i^2r_o^2}\tau,
\end{equation}
where $L_\mathrm{mid}$ is the length of the middle section of the inner cylinder, $\rho$ is the fluid density, and $\tau$ is the torque on the middle section of the inner cylinder.

The temperature is controlled by cooling the top and bottom lids. We measure the fluids temperature by three PT100 temperature sensors mounted flush on the inner cylinder, one in each part of the inner cylinder. The temperature is kept within \unit{1.5}{K} throughout the duration of the measurement and within \unit{0.1}{K} in space. We correct the viscosity and density of the fluid for the changes in temperature throughout the measurement. During the measurements we slowly accelerate the inner cylinder to increase the Taylor number. The inner cylinder is accelerated quasistatically such that the system is in equilibrium and that there is no significant contribution to the torque caused by the accelerating cylinders ($\tau_\mathrm{acc}=I\Dot{\omega}$).

Bubbles in the flow are generated by entrainment, hence no bubbles need to be injected. We fill the setup for 96\% with water and leave 4\% empty with regular air. Due to the very turbulent flow, air is very quickly entrained into the water and distributed vertically along the cylinder. The bubble sizes in the flow are therefore not affected by any initial effects of injection, but only dependent on the equilibrium of turbulent break-up and coalescence.

As fresh water we use decalcified water. In \cite{Blaauw_2023} we have analysed and presented the salt concentrations present in the base water. We add $\mathrm{MgCl}_2$, $\mathrm{Na}_2\mathrm{SO}_4$, and substitute sea salt to the fresh water for concentrations between \unit{0}{\gram/\kilogram} and \unit{35}{\gram/\kilogram}, where the latter is around the concentration of salt present in the ocean, in steps of \unit{8.75}{\gram/\kilogram}. The sea salt is prepared according to the ASTM D1141-98 standard and the contents are presented in table \ref{tab:SaltDR_seasalt}. Next to that, in a separate experiment, we also added $\mathrm{NaCH}_3\mathrm{COO}$ in a concentration between \unit{0}{\gram/\kilogram} and \unit{105}{\gram/\kilogram} in steps of \unit{35}{\gram/\kilogram}. Similar to \cite{Blaauw_2023} we define a normalised concentration $\Tilde{S}=\frac{S}{S_\mathrm{ocean}}$, where $S$ is the concentration of salt in g of salt per kg of solution and $S_\mathrm{ocean}=\unit{35}{\gram\per\kilogram}$ is the total concentration of salt found in seawater.

\begin{table}
    \centering
    \begin{tabular}{|c|c|c|c|}\hline
        Salt & Mass & Moles  & Molarity\\
         & [\%] & [\%] & [\milli\mole\per\liter]\\ \hline
        $\mathrm{NaCl}$ & 58.490 & 79.7 & 358\\
        $\mathrm{MgCl}_2\cdot6\mathrm{H}_2\mathrm{O}$ & 26.460 & 10.4 & 46.6 \\
        $\mathrm{Na}_2\mathrm{SO}_4$ & 9.750 & 5.47 & 24.6 \\
        $\mathrm{CaCl}_2$ & 2.765 & 1.98 & 8.91 \\
        $\mathrm{KCl}$ & 1.645 & 1.76 & 7.89\\
        $\mathrm{NaHCO}_3$ & 0.477 & 0.452 & 2.03\\
        $\mathrm{KBr}$ & 0.238 & 0.159 & 0.715 \\
        $\mathrm{H}_3\mathrm{BO}_3$ & 0.071 & 0.0929 & 0.418 \\
        $\mathrm{SrCl}_2\cdot6\mathrm{H}_2\mathrm{O}$ & 0.095 & 0.284 & 0.127 \\
        $\mathrm{NaF}$ & 0.007 & 0.0133 & 0.0596 \\\hline
    \end{tabular}
    \caption{Contents of the artificial sea salt (ASTM D1141-98 by Lake Products Company LLC) by weight percent. Molarities have been calculated assuming \unit{35}{\gram} dissolved salt per kilogram of solution.}
    \label{tab:SaltDR_seasalt}
\end{table}

Densities and viscosities of the working fluid are calculated using formula \ref{eq:SaltDR_saltviscdens}a,b respectively, where $\rho_{\mathrm{H}_2\mathrm{O}}$ and $\nu_{\mathrm{H}_2\mathrm{O}}$ are the density and kinematic viscosity of fresh water only dependent on the temperature $T$, and $K_\rho$ and $K_\nu$ are functions dependent on the type of salt and its concentration $S$. The functions $K_{\rho,\nu}$ are determined by measuring the density and viscosity of samples of the working fluid used during the experiments. The density is measured using an Anton Paar DMA 35 handheld density meter with an accuracy of $\unit{0.1}{\kilogram\per\meter^3}$. The viscosity is measured using an Ubbelohde viscometer with a capillary diameter of \unit{0.6}{\milli\meter} with an accuracy of $\unit{0.003}{\milli\meter^2\per\second}$. For experiments with bubbles the viscosity and density are corrected using the Einstein equations (\cite{einstein_1905}), see equation \ref{eq:SaltDR_einstein}.

\begin{subequations}
\begin{equation}
    \rho_l = K_\rho(S) \rho_{\mathrm{H}_2\mathrm{O}}(T)
    \label{eq:SaltDR_saltviscdens_dens}
\end{equation}
\begin{equation}
    \nu_l = K_\nu(S) \nu_{\mathrm{H}_2\mathrm{O}}(T)
    \label{eq:SaltDR_saltviscdens_visc}
\end{equation}
\label{eq:SaltDR_saltviscdens}
\end{subequations}

\begin{subequations}
\begin{equation}
    \rho = (1-\alpha)\rho_l
    \label{eq:SaltDR_einstein_dens}
\end{equation}
\begin{equation}
    \nu = (1+\frac{5}{2}\alpha)\nu_l
    \label{eq:SaltDR_einstein_visc}
\end{equation}
\label{eq:SaltDR_einstein}
\end{subequations}

Next to the setup we place a high speed camera (Photron mini AX200) to record the bubbles inside the flow, see figure \ref{fig:saltDR_setup}. The camera captures images at \unit{50}{fp\second} to capture statistically independent images as to determine bubble sizes. We focus on a region of approximately $\unit{2.5}{\centi\meter}\times\unit{2.5}{\centi\meter}$, close to the outer cylinder and at midheight. Note that we are unable to optically measure the bubbles closer to the inner cylinder as we are there unable to image the bubbles for all salt concentrations and Taylor numbers. Especially for larger Taylor numbers and high salt concentrations the fluid becomes optically opaque, as can be seen in figures \ref{fig:SaltDR_Images1} and \ref{fig:SaltDR_Images2}.

\section{Results}
\label{sec:SaltDr_Results}
During the measurements the torque on the inner cylinder is monitored and we use equation \ref{eq:SaltDR_Nusselt} to determine the Nusselt number. These measurements are compared to the single phase case, i.e. when there are no bubbles present in the flow, to determine the drag reduction:
\begin{equation}\label{eq:SaltDR_DR}
    \mathrm{DR(\%)}=\frac{\mathrm{Nu}_\omega[\mathrm{Ta},\alpha=0]-\mathrm{Nu}_\omega[\mathrm{Ta},\alpha=4\%]}{\mathrm{Nu}_\omega[\mathrm{Ta},\alpha=0]},
\end{equation}
where $\mathrm{Nu}_\omega[\mathrm{Ta},\alpha=0]$ is the Nusselt number as function of the Taylor number in the single phase case and $\mathrm{Nu}_\omega[\mathrm{Ta},\alpha=4\%]$ is the Nusselt number for a Taylor number in the bubbly case.

\begin{figure}
    \centering
    \includegraphics[width=0.8\linewidth]{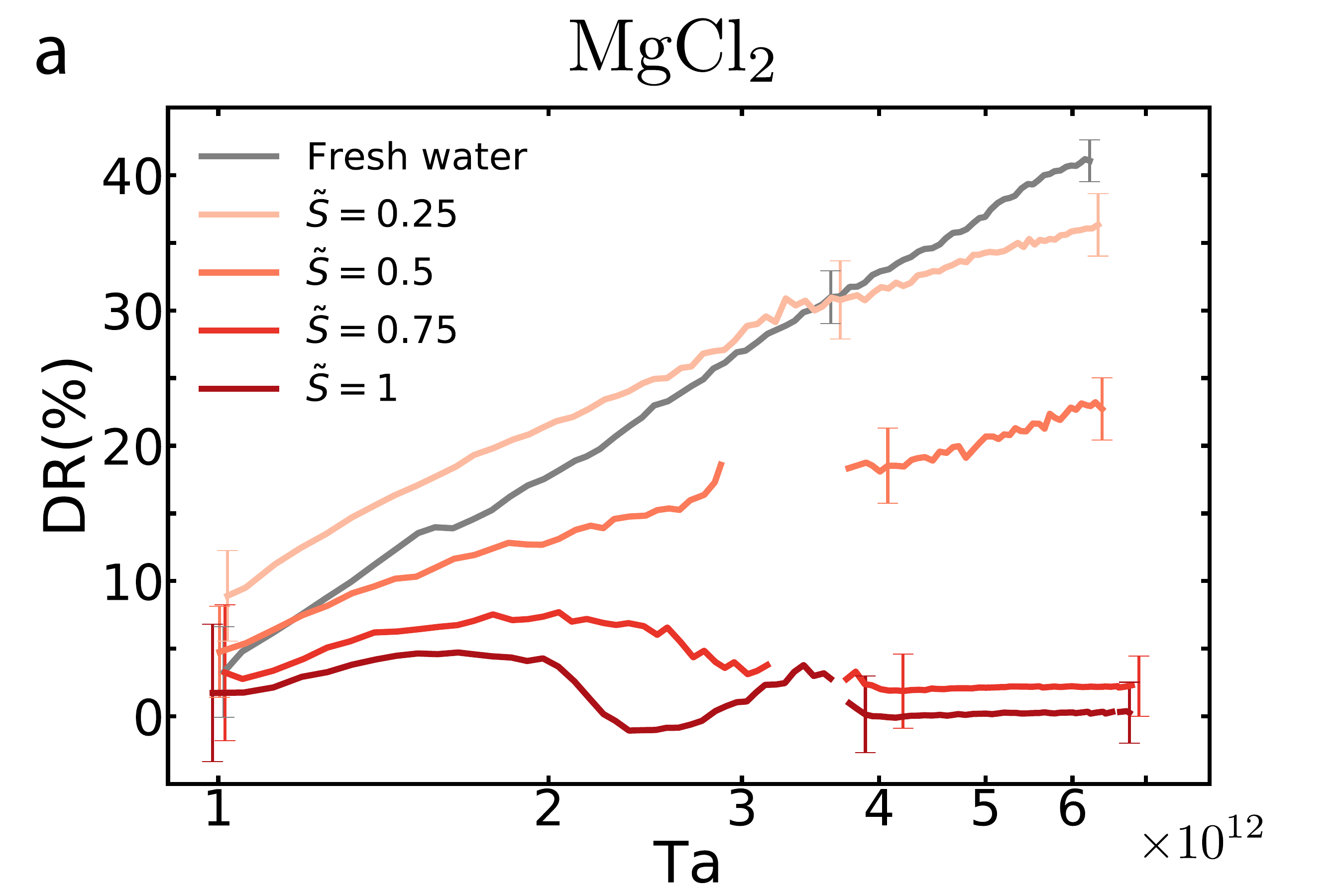}\\ \vspace{2pt}
    \includegraphics[width=0.8\linewidth]{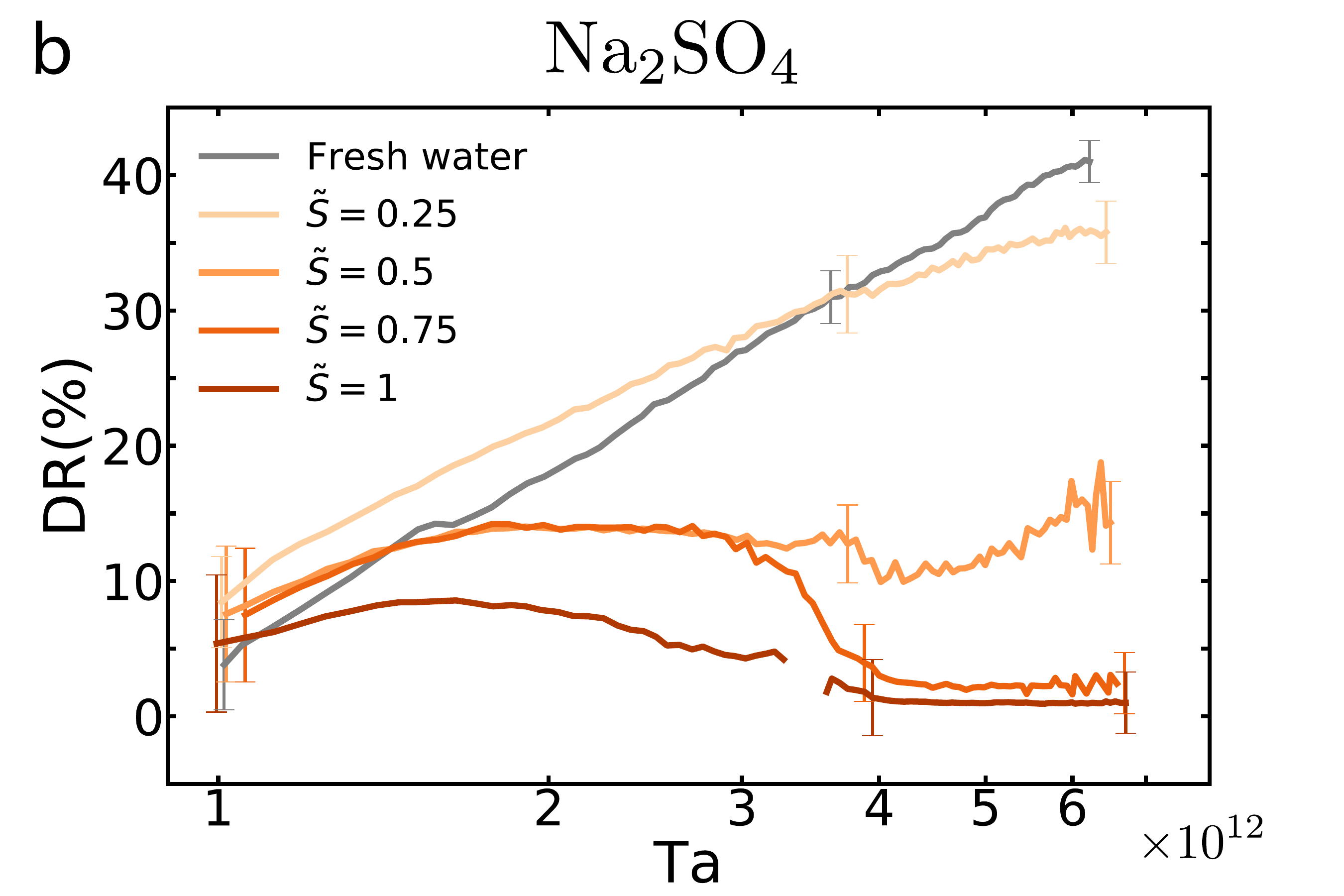}\\ \vspace{2pt}
    \includegraphics[width=0.8\linewidth]{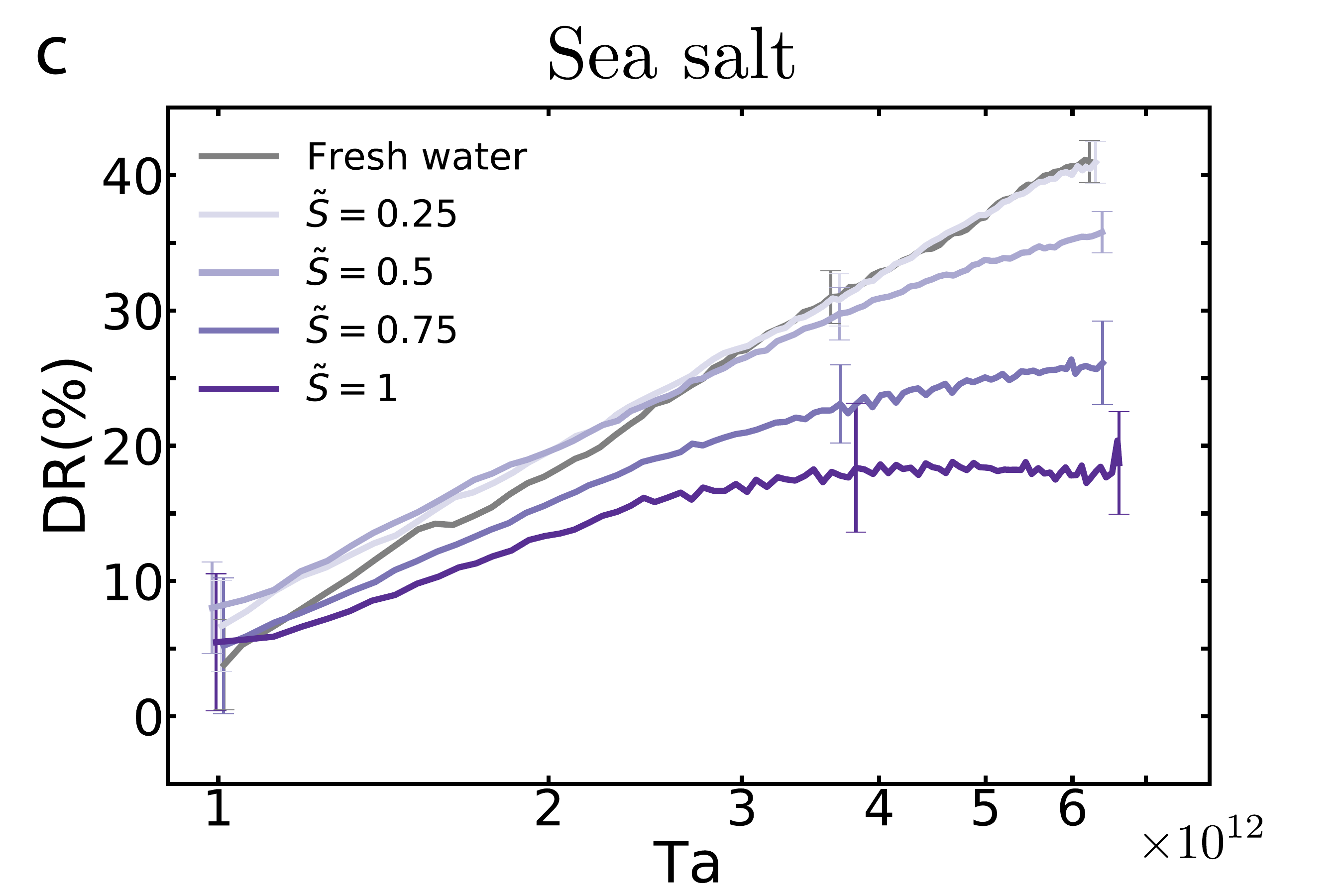}\\ \vspace{2pt}
    \includegraphics[width=0.8\linewidth]{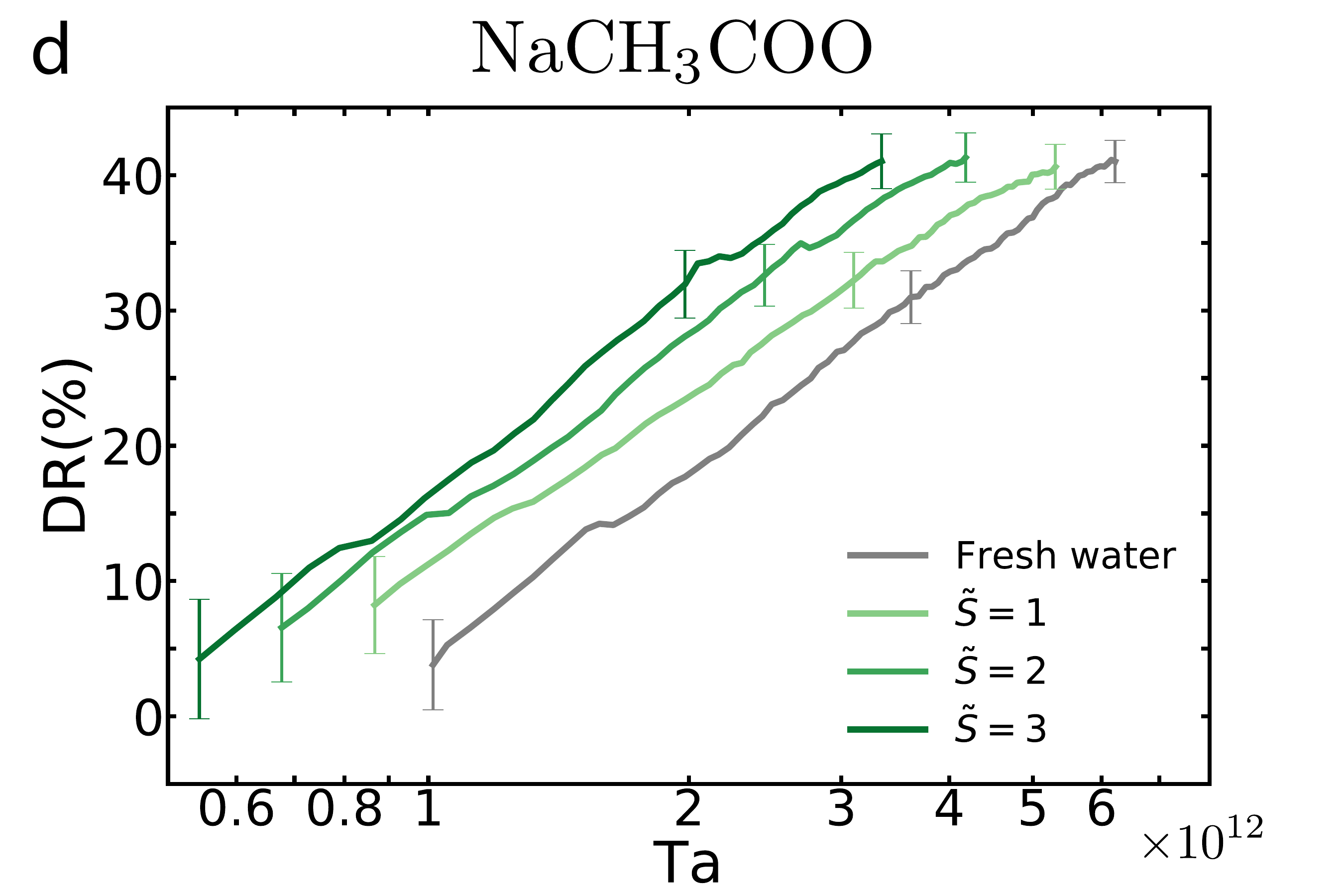}
    \caption{Drag reduction measured with a: magnesium chloride solutions, b: sodium sulfate solutions, c: substitute sea water, d: sodium acetate solutions. Gaps in the measurements in a and b are caused by unreliable torque measurements due to high vibrations in the setup.}
    \label{fig:saltDR_DragReduction}
\end{figure}

The drag reduction for different salt solutions is presented in figure \ref{fig:saltDR_DragReduction}, where all measurements are at least repeated once to ensure repeatability. Gaps in the data corresponds to regimes where due to vibrations (measured continuously during our experiments) the accuracy of our data can not be guaranteed as for certain regimes the bubbles are not uniformly distributed around the cylinder which can trigger eigenfrequencies of the mechanical system. The drag reduction for fresh water increases for increasing Taylor number up to a drag reduction of $\mathrm{DR}\approx40\%$ for the highest $\mathrm{Ta}$ tested, as was reported before (\cite{Verschoof_2016_surfactant,vGils_2011_T3C,Blaauw_2023}).

For a magnesium chloride solution, results are presented in figure \ref{fig:saltDR_DragReduction}a. For a low concentration, $\Tilde{S}=0.25$, the drag reduction is slightly higher than with fresh water for $\mathrm{Ta}\leq3.5\times10^{12}$, however this difference is within the experimental error bars. This slight increase in drag reduction might be caused by a better axial distribution for the smaller bubbles for $\Tilde{S}=0.25$ compared to fresh water. For the highest Taylor numbers tested the drag reduction with $\mathrm{MgCl_2}$ is slightly lower as compared to fresh water. Increasing the magnesium chloride concentration leads to a reduction of the drag reduction, i.e. a higher drag than the fresh water case with bubbles. For the highest magnesium chlorides concentrations investigated, $\Tilde{S}\geq0.75$, almost no drag reduction is observed and the highest two concentrations tested show no significant difference.

The drag reduction for sodium sulfate solutions in different concentrations, see figure \ref{fig:saltDR_DragReduction}b, look very similar to those of the magnesium chloride solutions. For a low concentration of sodium sulfate, $\Tilde{S}=0.25$, we observe a slight increase in the drag reduction for low Taylor numbers, $\mathrm{Ta}\leq3.5\times10^{12}$, however, this increase remains within the error margins of the measurement and is therefore not significant. For higher Taylor numbers, $\mathrm{Ta}\geq3.5\times10^{12}$, the drag reduction is slightly reduced. Increasing the sodium sulfate concentration decreases the drag reduction. For a concentration of $\Tilde{S}=0.75$ we observe some drag reduction of around $\mathrm{DR}\approx10\%$ for Taylor numbers $\mathrm{Ta}\leq3.5\times10^{12}$. For higher Taylor numbers the drag reduction reduces and almost disappears. For the highest sodium sulfate concentration tested almost no drag reduction is observed throughout the experiment, similar to the case of the highest magnesium chloride concentration.

For substitute sea water, see figure \ref{fig:saltDR_DragReduction}c, a low salt concentration $\Tilde{S}=0.25$ has no effect on the drag reduction over the investigated range of Taylor numbers. For $\mathrm{Ta}\leq2\times10^{12}$ the addition of substitute sea salt seems to only have a minor effect on the drag. Increasing the salt concentration leads to less effective drag reduction for higher Taylor numbers. For $\Tilde{S}\leq0.5$ the drag reduction always increases with increasing $\mathrm{Ta}$. For higher salt concentrations the increase seems to flatten and at $\Tilde{S}=1$ the drag reduction does not vary with $\mathrm{Ta}$ for $\mathrm{Ta}\geq3\times10^{12}$. At the salt concentration as found in sea water, $\Tilde{S}=1$, a drag reduction up to $\mathrm{DR}\approx20\%$ is still observed around $\mathrm{Ta}=6\times10^{12}$.

Sodium acetate is reported to not affect bubble coalescence (\cite{Craig_1993}), hence we would not expect an effect of salt concentration on the drag reduction, especially no reduction of the drag reduction. In fact, as seen in figure \ref{fig:saltDR_DragReduction}d, we find that the drag reduction even \emph{increases} when sodium acetate is added to the water. The drag reduction increases for increasing $\mathrm{Ta}$ for all concentrations considered up to a drag reduction of approximately $\mathrm{DR}=40\%$. Note that the different range of $\mathrm{Ta}$ investigated here as compared to the previous ones is due to the different viscosities for these salt concentrations.

\begin{figure}
    \centering
    \includegraphics[width=0.8\linewidth]{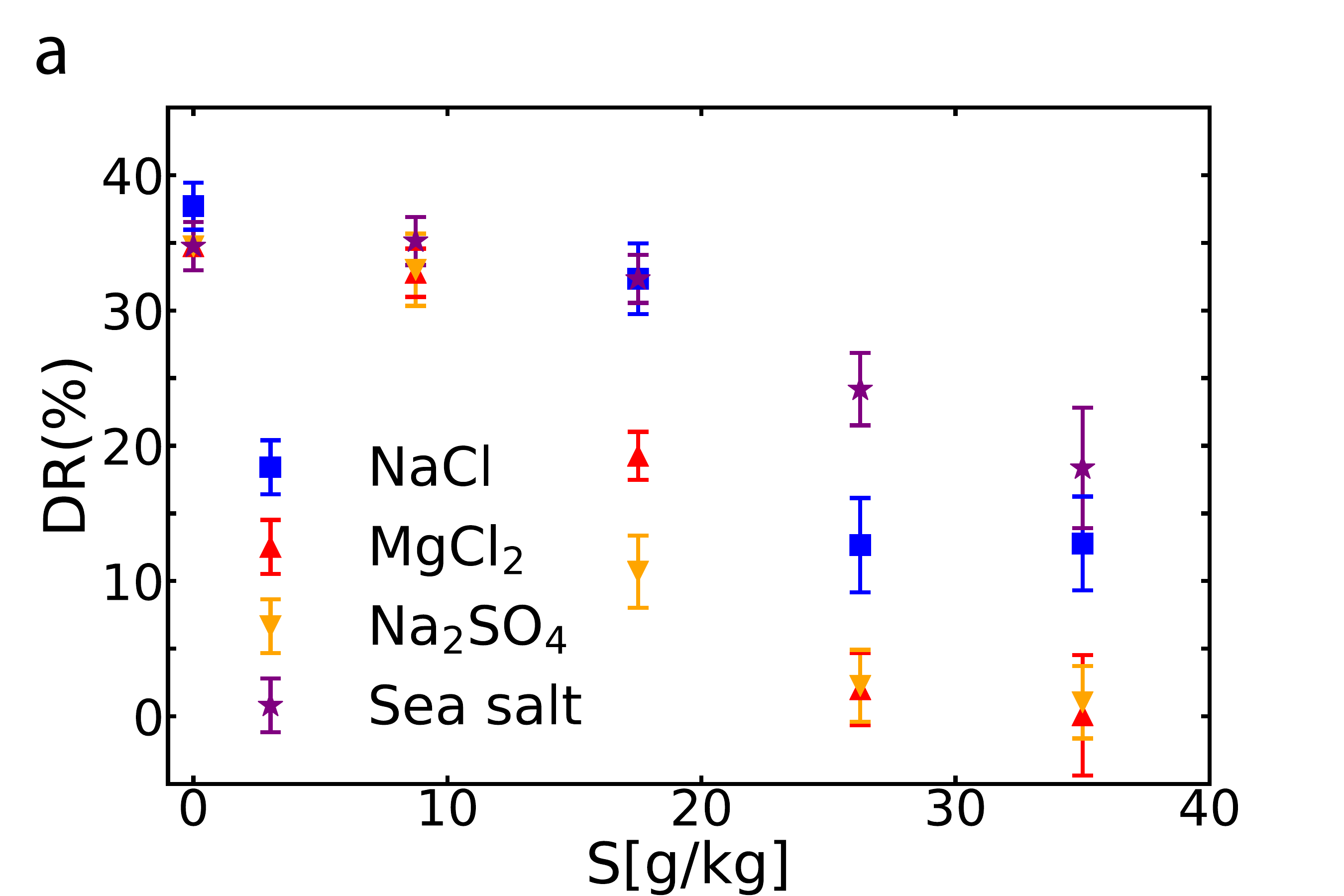}\\ \vspace{2pt}
    \includegraphics[width=0.8\linewidth]{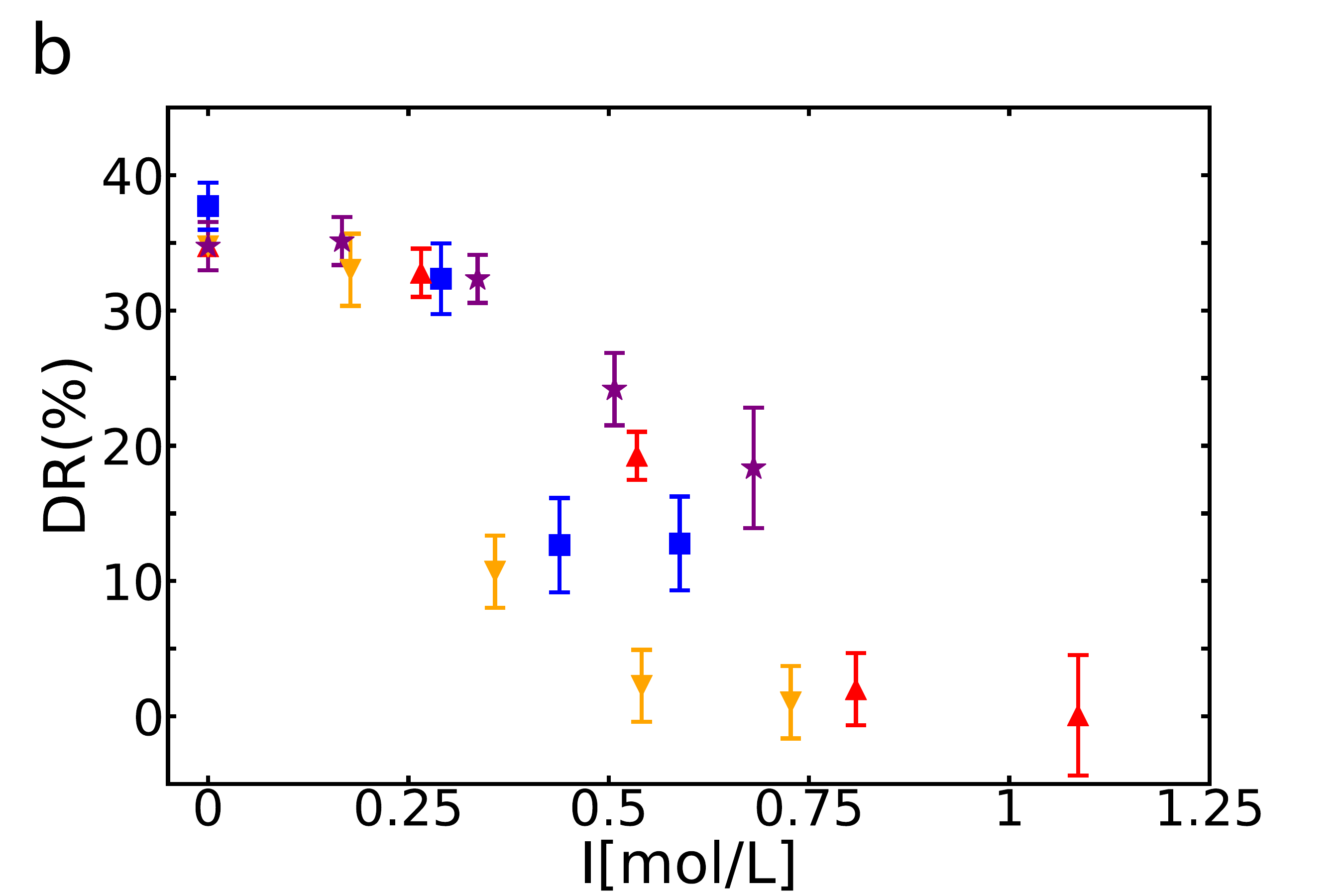}
    \caption{Comparison of drag reduction between different salt solutions by a: their concentration in g of salt per kilogram of solution and b: their ionic strength. For all cases the drag reduction is measured at $\mathrm{Ta}\approx4.5\times10^{12}$ and with an air volume fraction of $\alpha=4\%$. Results of sodium chloride, indicated by the blue squares, taken from \cite{Blaauw_2023}.}
    \label{fig:saltDR_SaltComparison}
\end{figure}

We compare the drag reduction for different salts in figure \ref{fig:saltDR_SaltComparison}. Here we only consider the salts that affect bubble coalescence and lead to a less effective drag reduction, i.e. in this figure we do not take $\mathrm{NaCH_3COO}$ into consideration. We also include the data from \cite{Blaauw_2023} of $\mathrm{NaCl}$ solutions. In \ref{fig:saltDR_SaltComparison}a we compare the different salt solutions using their concentration. The $\mathrm{MgCl}_2$ and $\mathrm{Na}_2\mathrm{SO}_4$ solutions show similar drag reduction for the same concentrations. Substitute seawater and $\mathrm{NaCl}$ show higher drag reductions compared to the other salts. This shows that just salt concentration is not a good indication of the effectiveness for drag reduction.

We calculate the ionic strength $I$ of the salt solutions using
\begin{equation}\label{eq:saltDR_IonicStrength}
    I = \frac{1}{2} \sum_{i=1}^{n}{c_iz^2_i},
\end{equation}
where $i$ sums over the number of ions, $c_i$ is the molar concentration of an ion and $z_i$ is its charge. For bubbles in salt solutions in flotations cells the bubble size collapses with the ionic strength (\cite{Quinn_2014,Sovechles_2015}). Comparing the different salts using the ionic strength gives a much better collapse of the drag reduction, however, definitely not a perfect one. We see that the drag reduction sharply decreases around an ionic strength $I\approx\unit{0.5}{mol\per\liter}$. For solutions with $I\geq\unit{0.7}{mol\per\liter}$ the drag reduction almost disappears. However, we only have data for $\mathrm{MgCl_2}$ and $\mathrm{Na_2SO_4}$ in this region and no datapoints for the other salts.
    \begin{figure}
        \centering
        \includegraphics[width=\linewidth]{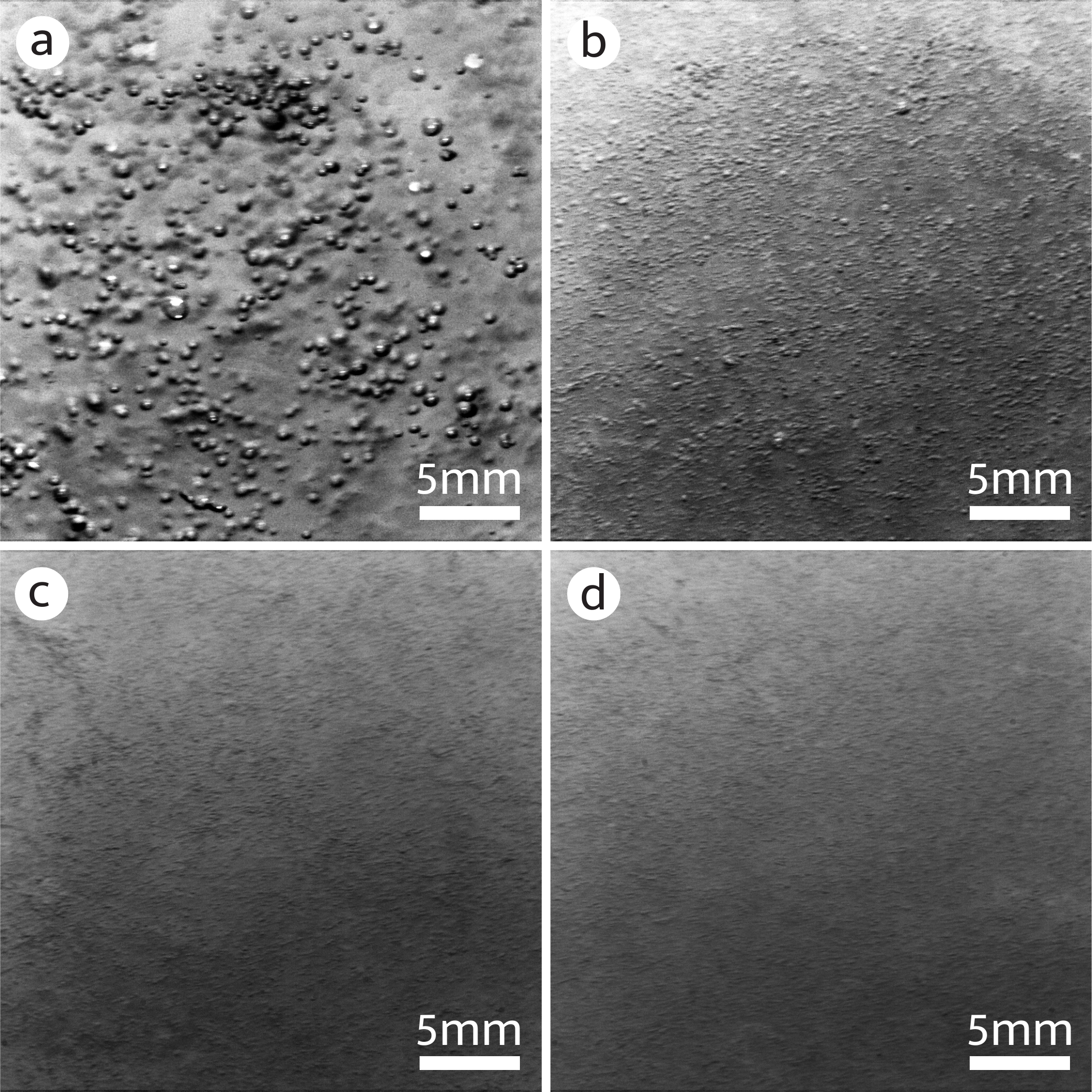}\\ \vspace{1pt}
        \includegraphics[width=\linewidth]{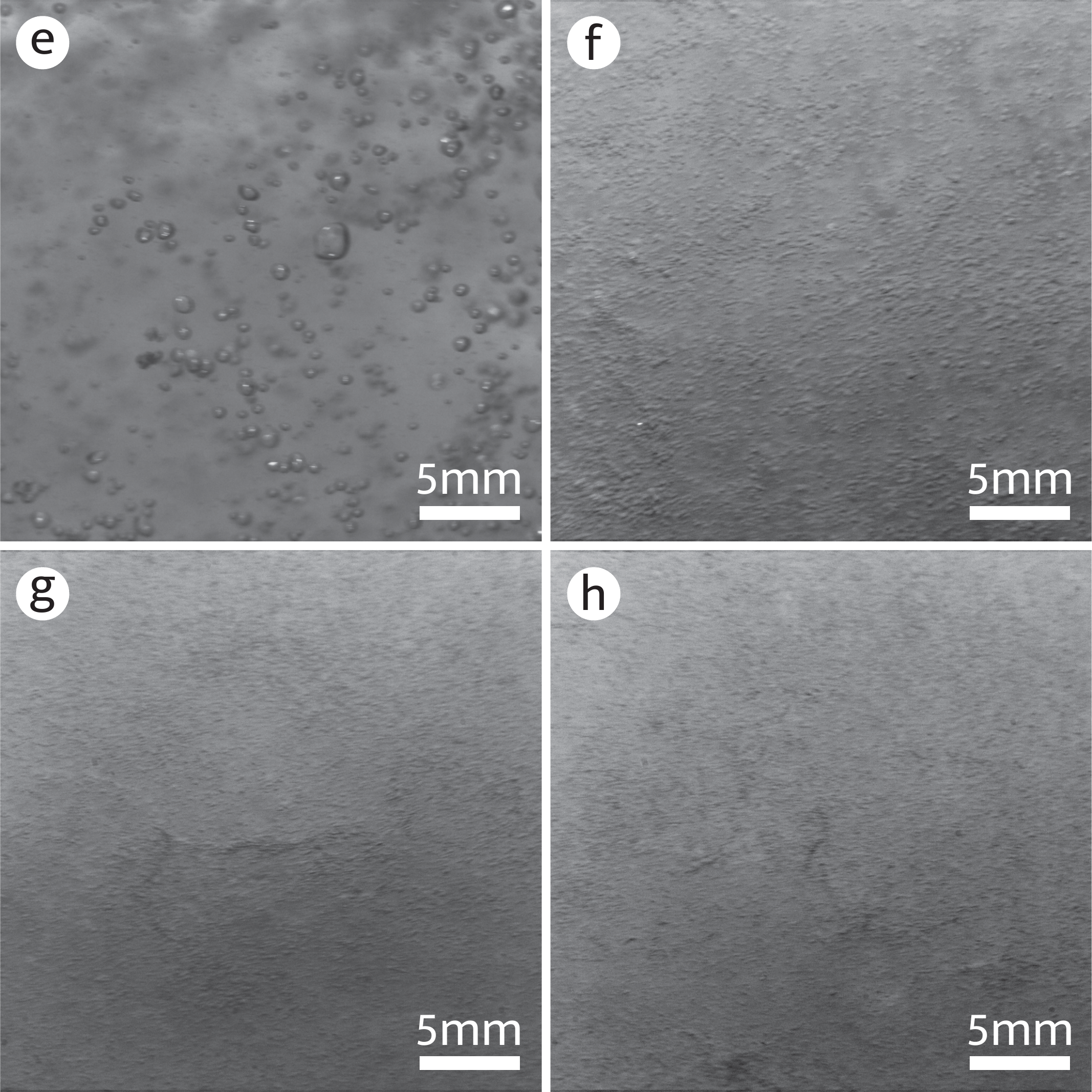}
        \caption{Typical images captured during experiments. Images taken at $\omega_i=2\pi\cdot\unit{20}{\hertz}$, hence $\mathrm{Ta}\approx6\times10^{12}$. Images taken at midheight and focal plane of the camera close to the outer cylinder. a--d: Magnesium chloride with a salt concentration of a: $\Tilde{S}=0.25$, b: $\Tilde{S}=0.5$, c: $\Tilde{S}=0.75$, d: $\Tilde{S}=1$, e--h: sodium sulfate with a salt concentration of e: $\Tilde{S}=0.25$, f: $\Tilde{S}=0.5$, g: $\Tilde{S}=0.75$, h: $\Tilde{S}=1$}
        \label{fig:SaltDR_Images1}
    \end{figure}
    \begin{figure}
        \centering
        \includegraphics[width=\linewidth]{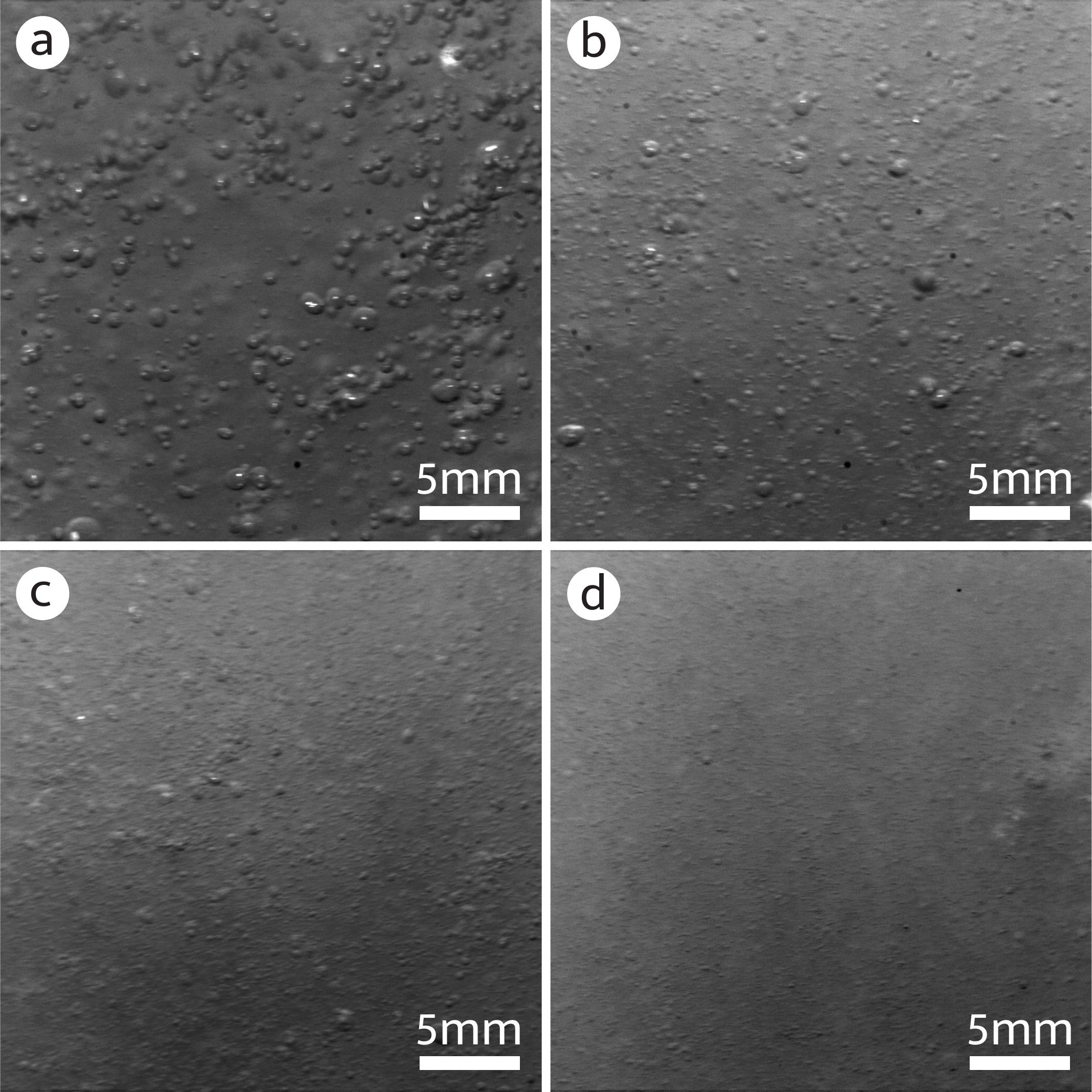}\\ \vspace{1pt}
        \includegraphics[width=\linewidth]{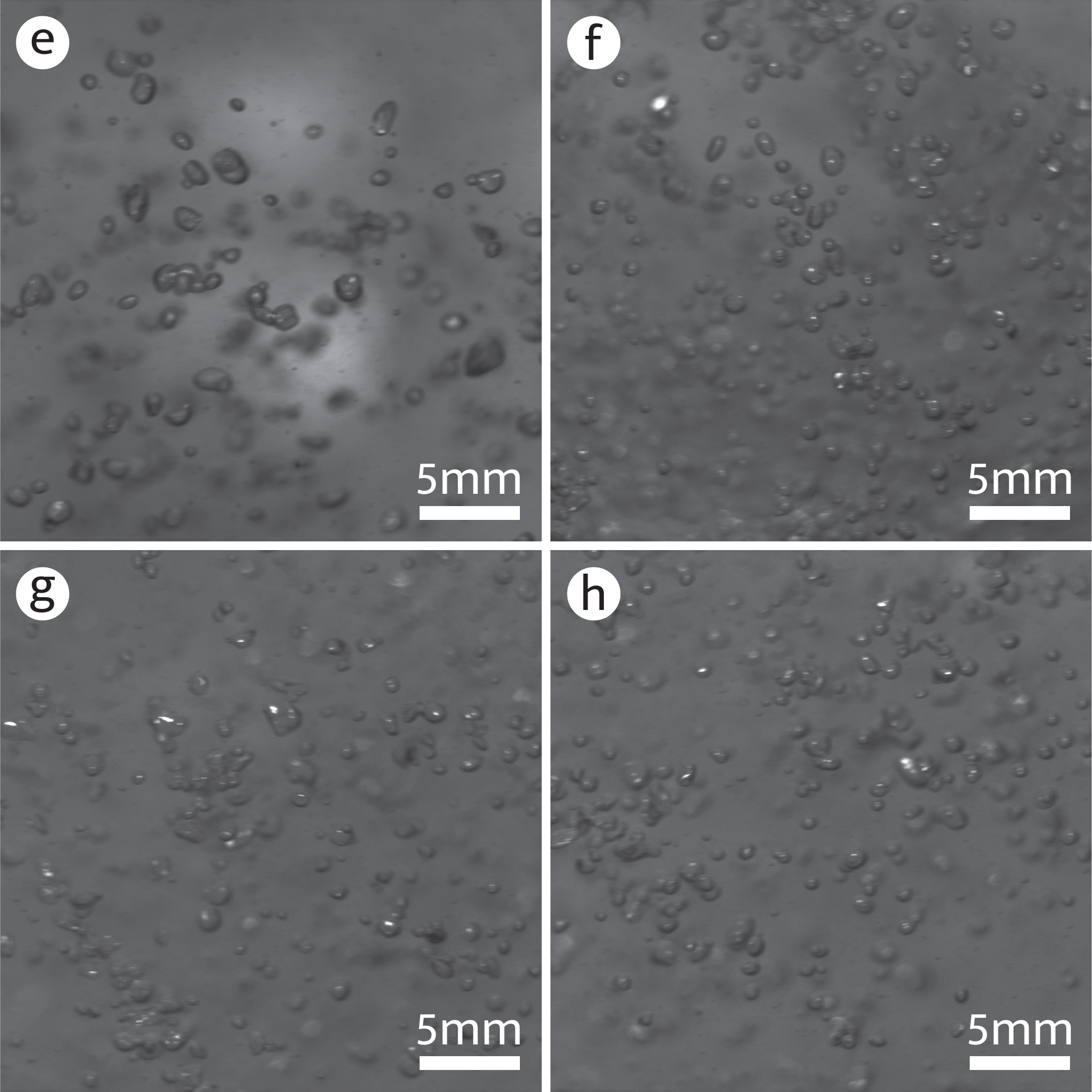}
        \caption{Typical images captured during experiments. Images taken at $\omega_i=2\pi\cdot\unit{20}{\hertz}$, hence at the largest $\mathrm{Ta}$ investigated for each case. Images taken at midheight and focal plane of the camera close to the outer cylinder. a--d: substitute seawater with a concentration of a: $\Tilde{S}=0.25$, b: $\Tilde{S}=0.5$, c: $\Tilde{S}=0.75$, d: $\Tilde{S}=1$, e: fresh water, f--h: sodium acetate with a concentration of f: $\Tilde{S}=1$, g: $\Tilde{S}=2$, h: $\Tilde{S}=3$.}
        \label{fig:SaltDR_Images2}
    \end{figure}

To elucidate the possible effect of coalescence and break up on bubbly drag reduction, we try to correlate bubble size and drag reduction. Typical images captured during the experiments are shown in figure \ref{fig:SaltDR_Images1} and figure \ref{fig:SaltDR_Images2}. These images are captured around $\mathrm{Ta}\approx6\times10^{12}$. For both $\mathrm{MgCl_2}$, figures \ref{fig:SaltDR_Images1}a--d, and $\mathrm{Na_2SO_4}$, figures \ref{fig:SaltDR_Images1}e--h, the bubbles become smaller with increasing salt concentration. At a salt concentration of $\Tilde{S}=0.25$ bubbles are already smaller compared to fresh water, figure \ref{fig:SaltDR_Images2}e, however still around $d_\text{bubble}\approx \unit{1}{\milli\meter}$ in size. For higher salt concentrations bubbles become very small with sizes below $d_\text{bubble}<\unit{1}{\milli\meter}$. Looking at the bubbles in the measurements with substitute seawater, figures \ref{fig:SaltDR_Images2}a--d, the bubbles are slightly larger than for $\mathrm{MgCl_2}$ and $\mathrm{Na_2SO_4}$. Especially for $\Tilde{S}=0.5$, figure \ref{fig:SaltDR_Images2}b, and $\Tilde{S}=0.75$, figure \ref{fig:SaltDR_Images2}c, we observe some more bubbles where $d_\text{bubble} = \mathcal{O}(\unit{1}{\milli\meter})$ for substitute seawater compared to $\mathrm{MgCl_2}$ and $\mathrm{Na_2SO_4}$ at the same concentrations. We note that our bubbling sizing for fresh water is compatible with the findings of \cite{vGils_2013_bubbledeform}, where they furthermore found that the bubble size is nearly independent on the radial position, see their figure 10.

The link between $\text{We}$ and drag reduction agree with earlier observations of \cite{Verschoof_2016_surfactant,vGils_2013_bubbledeform,Blaauw_2023}, where the importance of the bubble Weber number was found, describing the ratio of the inertia of the surrounding liquid and the surface tension of the bubble:

\begin{equation}\label{eq:saltDR_We}
    \mathrm{We} = \frac{\rho_l u^{\prime 2} d_\text{bubble}}{\sigma},
\end{equation}

where $u'$ are the velocity fluctuations and $\sigma$ is the surface tension. The surface tension is determined using the data from \cite{Dutcher_2010_surfacetension} for NaCl, $\mathrm{MgCl_2}$, and $\mathrm{Na_2SO_4}$ solutions and using data from \cite{Minofar_2007_surftensNaAc} for $\mathrm{NaCH_3COO}$ solutions. Since adding salt only has a minor effect on the surface tension (e.g.\  $\sigma(\tilde S=0)=\unit{72.8}{\milli\newton\per\meter}$ and $\sigma(\tilde S=1)=\unit{73.7}{\milli\newton\per\meter}$ for the case of $\mathrm{NaCl}$) and because we do not have an exact measurement of the surface tension of seawater, we approximated the surface tension of seawater by the surface tension of NaCl solutions with the same salt concentration. The bubble Weber number characterises the forces of the velocity fluctuations with surface tension forces, where bubbles with $\mathrm{We}>1$ are deformed by the turbulent flow and bubbles with $\mathrm{We}<1$ remain spherical. We hypothesize that small, and therefore rigid, bubbles can effectively transfer momentum as their interface is stiff, whereas large, deformable bubbles are unable to transfer momentum, as their interface deforms and thus both sides of the bubbles are only weakly coupled.
Looking at images taken for $\mathrm{NaCH_3COO}$, figures \ref{fig:SaltDR_Images2}f--h, we do not observe a significant change in bubble sizes. Since $\mathrm{NaCH_3COO}$ does not inhibit bubble coalescence the equilibrium of bubble coalescence and turbulent break-up does not alter significantly, the bubbles remain comparable in size as compared to fresh water.

\begin{figure}
    \centering
    \includegraphics[width=\linewidth]{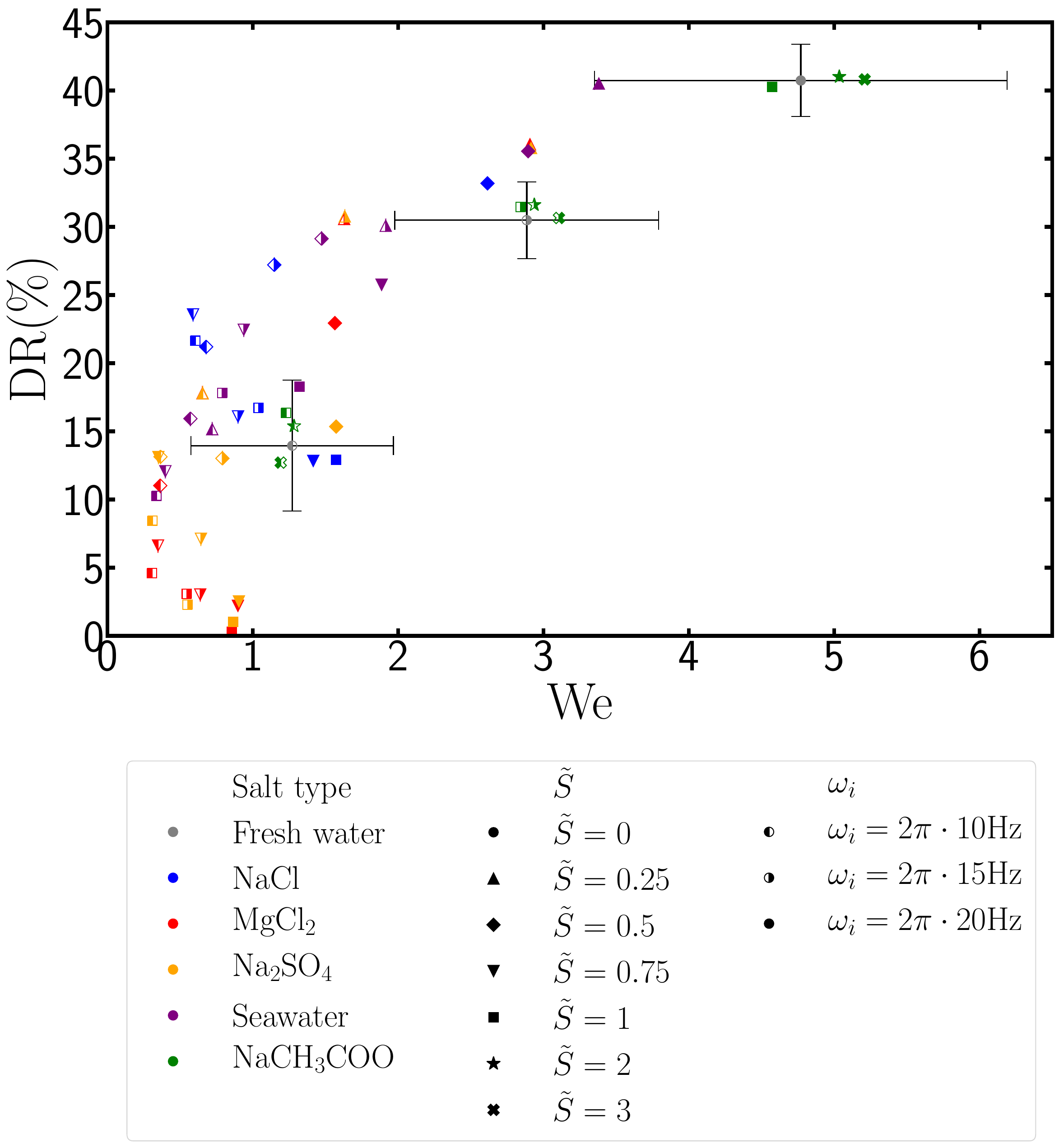}
    \caption{Observed drag reduction with the estimated bubble Weber number. Colours correspond to the colours used in figure \ref{fig:saltDR_DragReduction}. Left hand side filled symbols correspond to measurements at $\omega_i=2\pi\cdot\unit{10}{\hertz}$, right hand side filled symbols to measurements at $\omega_i=2\pi\cdot\unit{15}{\hertz}$, and closed symbols to measurements at $\omega_i = 2\pi\cdot\unit{20}{\hertz}$. The horizontal interval markers indicate the spread ($\pm1$ standard deviation obtained from $\mathcal{O}(200)$ bubbles) in the bubble size distribution for the Weber number and the vertical interval markers indicate the uncertainty in the drag measurements. Data for $\mathrm{NaCl}$ from measurements in \cite{Blaauw_2023}. We find that a Pearson correlation coefficient between $\text{DR}$ and $\text{We}$ (combining all salts, salinities, and Taylor numbers) yields $\rho_{\text{We},\text{DR}} = 0.83$.}
    \label{fig:SaltDR_DRWe}
\end{figure}

This, however, does not explain the increase in drag reduction for the same $\mathrm{Ta}$ with increasing $\mathrm{NaCH_3COO}$ concentration. Hence, we estimated the bubble Weber numbers for all salts used and relate this to the drag reduction, see figure \ref{fig:SaltDR_DRWe}. The bubble sizes are obtained using the images of figure \ref{fig:SaltDR_Images1} and figure \ref{fig:SaltDR_Images2}. The Weber number is calculated using equation \ref{eq:saltDR_We}. Similar as we did before in \cite{Blaauw_2023}, the velocity fluctuations are estimated at $u^\prime = 0.025 u_i$ based on the measurements of \cite{vGils_2013_bubbledeform}, where $u_i$ is the velocity of the inner cylinder. These measurements have been performed in the same experimental facility as our measurements, around $\text{Re}_i = 1\times10^6$ using a similar air volume fraction as our measurements. Though these measurements have been taken using fresh water, hence with large bubbles in the flow, the velocity fluctuations in the salt water cases are expected to be of a similar order of magnitude. We estimated the bubble Weber number at inner cylinder velocities of $\omega_i = 2\pi\cdot\unit{10}{\hertz}$, $\omega_i = 2\pi\cdot\unit{15}{\hertz}$, and $\omega_i = 2\pi\cdot\unit{20}{\hertz}$. These points correspond approximately to the places where the error bars are drawn in figure \ref{fig:saltDR_DragReduction}. We see from figure \ref{fig:SaltDR_DRWe} that there is a good correlation between the bubble Weber number and the observed drag reduction, and the Pearson correlation coefficient is found to be $\rho_{\text{We},\text{DR}}  = 0.83$. The horizontal interval markers indicate the spread in the bubble size distribution for the Weber number and the vertical interval markers indicate the uncertainty in the drag measurements. Flows with high Weber number bubbles are effective in reducing the drag on the inner cylinder, where for flows with bubbles with low Weber number the drag reduction in minimal. This stresses the importance of bubble deformability for effective drag reduction. At low Weber numbers there is some more spread in the plot. This is possibly due to the larger relative errors in determining the Weber number of small bubbles. Also, it is likely that other proposed mechanisms for bubbly drag reduction, for example micro-bubbles pushing vortical structures away from the wall as proposed by \cite{ferrante_elghobashi_2004}, might become more dominant in this regime, leading to more spread in the data.

\newpage
\section{Conclusion}
\label{sec:SaltDR_conclusion}
We experimentally investigated the effects of different salts on bubbly drag reduction in turbulent Taylor--Couette turbulence. As an extension to our previous work (\cite{Blaauw_2023}) we used $\mathrm{MgCl_2}$, $\mathrm{Na_2SO_4}$, and substitute seawater solutions with salt concentrations ranging from fresh water ($\Tilde{S}=0$) to equivalent salt concentrations as found in seawater ($\Tilde{S}=1$). Next to these salts that inhibit bubble coalescence we also investigated $\mathrm{NaCH_3COO}$ solutions, a salt which does not inhibit bubble coalescence, in concentrations ranging from fresh water ($\Tilde{S}=0$) to 3 times the salt concentration found in seawater ($\Tilde{S}=3$).

All measurements were performed with a global air volume fraction of $\alpha=4\%$. This leads in fresh water to a drag reduction of $40\%$ at $\mathrm{Ta}=6\times10^{12}$. Adding salts that inhibit bubble coalescence lead to smaller bubbles in the flow. These bubbles are less effective for drag reduction. For the highest concentrations tested in $\mathrm{MgCl_2}$ and $\mathrm{Na_2SO_4}$ solutions the drag reduction almost completely disappears. For the same concentration of artificial sea salt we still observe a drag reduction of 20\%. We compare the drag reductions found for different salt solutions around $\mathrm{Ta}\approx4.5\times10^{12}$. We find that the ionic strength gives a reasonable indication of the observed drag reduction for the tested salts, where for $I\geq\unit{0.7}{mol\per\liter}$ almost no drag reduction is observed. This indication only works for salts that inhibit bubble coalescence and not for salts that do not affect bubble coalescence. Also, whether this indication works for other salts inhibiting coalescence but not tested here remains an open question.

Using a salt solution that has no effect on bubble coalescence we see no clear difference in the bubble size distribution between salt water and fresh water. The drag reduction at a specific Taylor number, however, is increased for an increasing salt concentration. To explain this we estimated the bubble Weber numbers for all salts tested using images obtained with a high speed camera. We relate the bubble Weber number to the observed drag reduction, which shows that for high drag reductions we need high bubble Weber numbers. This stresses the importance of bubble deformability for effective bubbly drag reduction.

Our finding provide new challenges to face for shipbuilders designing ships with fuel saving techniques for the sea and ocean. Moreover, our findings can explain why bubbly drag reduction in rivers, with fresh water, show larger drag reduction as compared to what has been seen in full scale tests in the ocean.

\section*{Acknowledgements}
This work has been supported by NWO through the AQUA program. We thank Gert-Wim Bruggert, Thomas Zijlstra, Martin Bos, Geert Mentink, and Jochem Paalman for technical support.

\end{document}